\documentclass[12pt]{article}
\usepackage{graphicx,amsmath}

\newcommand\pubnumber{SNSN-323-63}
\newcommand\pubdate{\today}

\def\institute{Departimento di Fisica ``E. Pancini'',\\
Universit\`a di Napoli Federico II and INFN, sezione di Napoli,\\
via Cintia I-80126 Napoli, Italy}

\def\Title#1{\begin{center} {\Large #1 } \end{center}}
\def\Author#1{\begin{center}{ \sc #1} \end{center}}
\def\Address#1{\begin{center}{ \it #1} \end{center}}

\newcommand\pubblock{\rightline{\begin{tabular}{l} \pubnumber\\
         \pubdate  \end{tabular}}}
\newenvironment{Abstract}{\begin{quotation}  }{\end{quotation}}
\newenvironment{Presented}{\begin{quotation} \begin{center} 
             PRESENTED AT\end{center}\bigskip 
      \begin{center}\begin{large}}{\end{large}\end{center} \end{quotation}}

\def\beq{\begin{equation}}
\def\eeq#1{\label{#1}\end{equation}}
\def\eeqn{\end{equation}}

\def\beqa{\begin{eqnarray}}
\def\eeqa#1{\label{#1}\end{eqnarray}}
\def\eeqan{\end{eqnarray}}

\let\bar=\overbar

\def\Dslash{\not{\hbox{\kern-4pt $D$}}}
\def\dslash{\not{\hbox{\kern-2pt $\del$}}}

\def\msb{{\bar{\ssstyle M \kern -1pt S}}}

\begin{document}
\begin{titlepage}
\pubblock

\vfill
\Title{Single top theory}
\vfill
\Author{ Francesco Tramontano} 
\Address{\institute}
\vfill
\begin{Abstract}
I breafly discuss very recent progress in the theoretical description of Standard Model
single top production at hadron colliders.
\end{Abstract}
\vfill
\begin{Presented}
$9^{th}$ International Workshop on Top Quark Physics\\
Olomouc, Czech Republic,  September 19--23, 2016
\end{Presented}
\vfill
\end{titlepage}
\def\thefootnote{\fnsymbol{footnote}}
\setcounter{footnote}{0}

\section{Introduction}
Many things have changed since the work of  Willenbrock and Dicus in 1986~\cite{Willenbrock:1986cr}
that showed the relevance of single top physics at hadron colliders using tree level diagrams.
In this presentation I'll review the progress of the last year, since the end of TOP2015.
Furthermore, I'll just present a selection of results, I apologise, there has been indeed a lot of progress
recently. The reader will note that ``single top'' production in hadron collisions is more and more an important
tool that triggers progress in high energy physics.
In the next three subsections I'll report about new computations for t-channel, tW associated production
and s-channel respectively.

\section{t-channel}

\begin{figure}[htb]
\centering
\includegraphics[height=7cm]{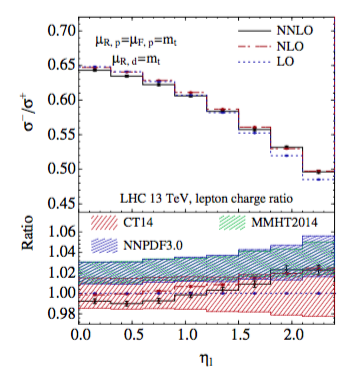}
\caption{Ratios of fiducial cross sections of top anti-quark to top quark production with decay
at 13 TeV as a function of the pseudorapidity of the charged lepton.}
\label{fig:1}
\end{figure}

In ref.~\cite{Berger:2016oht} the authors report about the computation of the Next-to-Next-to Leading Order (NNLO)
QCD corrections to the t-channel single top production and decay. The computation has been performed
using the structure function approach neglecting the color transfer among the two fermion lines, furthermore, the 
merging of the corrections for the production and decay of the top quark has been worked out in the narrow width approximation.
From this differential computation one expects the highest precision on distributions that are inclusive or insensitive to
the invariant mass of the reconstructed top, like the light jet transverse momentum and pseudo-rapidity.
Furthermore, such computations might be relevant for pdf studies.
Adopting the following definition for the NNLO cross section:
\begin{eqnarray}
\delta \sigma^{NNLO}&=&\frac{1}{\Gamma^0_t}\,\left[ {\rm d}\sigma^2 {\rm d}\Gamma^0_t + {\rm d}
\sigma^1\,\left({\rm d} \Gamma^1_t -\frac{\Gamma^1_t}{\Gamma^0_t}\,{\rm d} \Gamma^0_t \right) \right.\\ \nonumber
&+& \left. {\rm d} \sigma^0\,\left( {\rm d} \Gamma^2_t - \frac{\Gamma^2_t}{\Gamma^0_t}\,d \Gamma^0_t
- \frac{\Gamma^1_t}{\Gamma^0_t}\,\left( d \Gamma^1_t -\frac{\Gamma^1_t}{\Gamma^0_t}\,{\rm d} \Gamma^0_t \right)\right) \right]
\end{eqnarray}
the results for the total cross section at LHC at 13TeV obtained using CT14 pdf and using $m_t=173.2$ as central scale
are reported in Table~\ref{tab:1}.
\begin{table}[htb]
\begin{center}
\begin{tabular}{l|c|c|c}  
inclusive [pb] 	&  LO 	&  NLO 	&  NNLO  		\\ \hline
$t$ quark  	&  $143.7^{+8.1\%}_{-10\%}$   	&  $138.0^{+2.9\%}_{-1.7\%}$  	&  $134.3^{+1.0\%}_{-0.5\%}$  	\\
$\bar{t}$ quark 	&  $  85.8^{+8.3\%}_{-10\%}$   	&  $  81.8^{+3.0\%}_{-1.6\%}$	&  $  79.3^{+1.0\%}_{-0.6\%}$ 	\\ \hline
\end{tabular}
\caption{Inclusive cross sections for top (anti-)quark pro- duction at 13 TeV at various orders in QCD.}
\label{tab:1}
\end{center}
\end{table}
One can observe the nice features of the higher order corrections that give a small
contribution and reduce the theoretical error.
The impact of higher order corrections on the lepton charge ratio is shown in Figure~\ref{fig:1}.

Another important step forward has been done improving the matching of parton shower to matrix elements that involve
a top-quark resonance.
The problem of the matching of the Next to Leading Order (NLO) computation to a Parton Shower (PS) in presence of internal
resonances is of prominent importance for the whole physics program at the LHC. Leading collaborations have started to
address this problem studying t-channel single top production.

With the general subtraction schemes like Catani and Seymour or Frixione, Kunszt and Signer, only the ingoing
and outgoing particle momenta are relevant and the counter events have the primary condition to adsorb the momenta
of the extra radiated parton among the particles of the Born process, irrespective of the presence of internal resonances.
This situation deteriorates the efficiency of the infrared cancellation.
In a NLO+PS computation in presence of resonances there are even more severe problems. In the case of the POWHEG
method for example, to work correctly, the ratio among the real emission and underlining Born matrix elements
must become large only in the collinear limit, while, if there are resonances, the general mappings that are employed
to attach the radiation can produce unphysical distortions in the Sudakov factor.
Last, but not less important, PS has to be instructed about when preserve resonance mass during the shower to
avoid arbitrary shifts of the resonance invariant mass resulting in an unphysical distortion of the top line shape.

Solutions to all the problems listed above have been proposed by two groups. The solution found in POWHEG
in~\cite{Jezo:2015aia} is based on a division of the contributions to the cross section following all the possible resonance
histories obtained introducing appropriate multiplicative factors for all the contributions. The other fundamental step is the introduction of
resonance aware mappings, in such a way to minimise the mismatch among real and subtractions.
A crucial aspect here is the delicate separation of the real contribution into resonance histories that preserve
collinear factorisation and allow for subtractions computed in the resonance frame.

Similarly, the solution found in MG5\_aMC@NLO~\cite{Frederix:2016rdc}
is based on the division of the contribution following the possible resonance histories performed combining input from
the amplitude structure, the FKS singular region and information on the kinematical configuration.
Then a remapping of the event kinematics is performed to match the resonance invariant mass, resorting on numerical
methods to compute the new Jacobian factor so that there is no need of further analytic integrations.

\begin{figure}[htb]
\centering
\includegraphics[height=5cm]{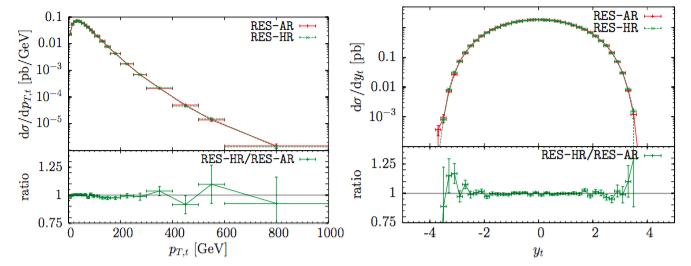}
\caption{$p_t$ (left) and rapidity (right) distributions of the top quark produced in 8TeV proton-proton collisions
generated with POWHEG-BOX-VRES.}
\label{fig:2}
\end{figure}

\begin{figure}[htb]
\centering
\includegraphics[height=5cm]{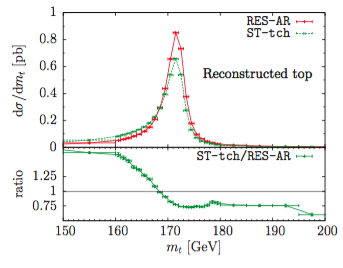}
\includegraphics[height=5cm]{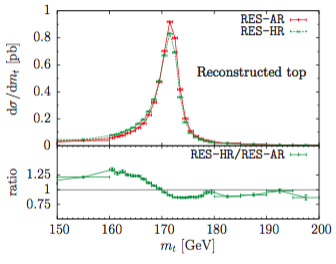}
\caption{Comparisons of the reconstructed top quark mass obtained with different generators
for 8TeV proton-proton collisions generated with POWHEG-BOX-VRES.}
\label{fig:3}
\end{figure}

The two collaborations have independently compared their different generators for the same process to understand the
effects of the new treatment of the radiation. 
In POWHEG-BOX-RES the authors simulated pp collisions at 8TeV, using the MSTW2008NLO pdf set
and assuming the following jet cuts: $p_T>25$, $|\eta|<4.5$ adopting the kt-algorithm with R=0.5.
The top quark distributions are given in Figure~\ref{fig:2}. They are obtained writing the Hardest Radiation (RES-HR ) of the event
or both the hardest radiation from the production and the one from the decay (RES-AR).
One observe good agreement among the two.
In Figure~\ref{fig:3} there is the reconstructed top mass distributions. On the left the All Radiation generation (RES-AR) is compared
to the one with hard radiation generated only in the production process and the top decay described by the PS.
A shift of about 1GeV is found. On the right side the AR and HR generators are compared and a less pronounced
difference is found for the top line shape.

With MG5\_aMC@NLO the authors had the same setup plus a cut on the reconstructed top mass
${\rm 140GeV<M(W^+,J_b)<200GeV}$.
In their analysis, the authors made a detailed study of the dependance of the predictions on the variation of the
technical parameters that in turn is useful to understand the remaining associated uncertainties.

\begin{figure}[htb]
\centering
\includegraphics[height=10cm]{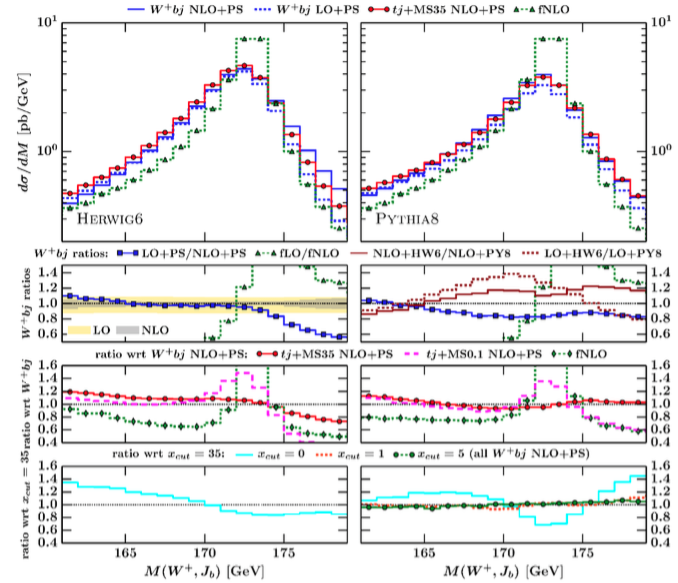}
\caption{invariant mass distribution of the the reconstructed top quark generated with MG5\_aMC@NLO.}
\label{fig:4}
\end{figure}

\begin{figure}[htb]
\centering
\includegraphics[height=10cm]{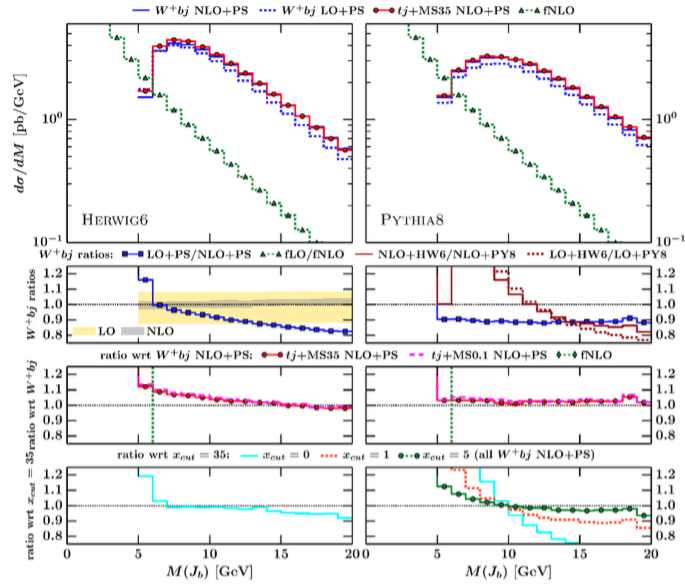}
\caption{Mass distribution of the primary b-jet in single top events generated with MG5\_aMC@NLO.}
\label{fig:5}
\end{figure}

In Figure~\ref{fig:4} and~\ref{fig:5} the invariant mass of the the reconstructed top quark and the mass of the 
primary b-jet are shown respectively. One observe how the PS smears and flatten the sharp peak of fixed order prediction
in the first and that the b-jet mass results much harder in NLO+PS wrt fixed order NLO.
The reader can find many distributions and details in the original works.

\section{tW associated production}

As for the tW associated production, recently the POWHEG-BOX-VRES collaboration
applied the new method for the treatment of the resonances histories to the computation
of the $b \bar{b}$ plus four leptons production at the LHC, and so completing the description
of $t \bar{t}$ pair production and decay at NLO+PS including all the non resonant and off shell contributions that
produce the same final state~\cite{Jezo:2016ujg}. This new generator is also an ideal generator for tW associated
production where all the effects of the interference with the $t\bar{t}$ production are consistently
included.

\begin{figure}[htb]
\centering
\includegraphics[height=6cm]{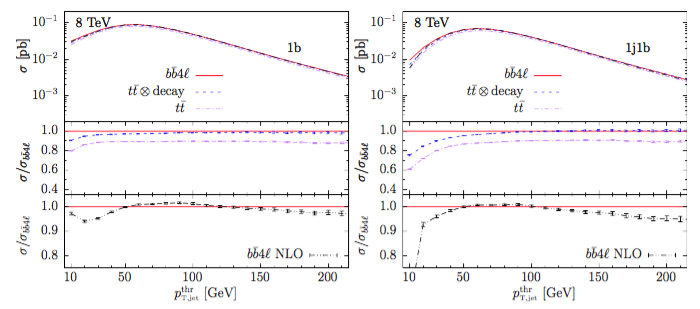}
\caption{Cross sections for exactly 1-bjet above a pt threshold, inclusive in the left panel and exclusive in the right panel,
produced with the bb4l POWHEG-BOX-VRES generator.}
\label{fig:6}
\end{figure}

In Figure~\ref{fig:6} are shown the cross sections for exactly 1-bjet above a $p_t$ threshold, inclusive in the left
panel and exclusive in the right panel. One can note an increasingly important single top
contribution at small $p_t$ jet threshold. A detailed study of Wt and HWt associated production
at NLO+PS has also been presented in~\cite{Demartin:2016axk}.

\section{s-channel}

Based on the results for the soft gluon resummation of Kidonakis~\cite{2006,2007,2010} in ref.~\cite{Alekhin:2016jjz}
Alekhin et al. implemented an approximated NNLO s-channel single top cross section. The authors then extracted the value
of the top mass from single top cross sections and $t\bar{t}$ production fitting the best experimental
determinations. Note that for the s-channel, Tevatron results are the ones with the highest precision.
The exercise has been performed for three different pdf sets and shows that slightly more consistent values
of the top mass are obtained with the ABM sets. Note however that the experimental cross sections determinations
are correlated to the assumptions on the top mass value and that this correlation is not available for all
the measurements and has not been taken into account in the work of ref.~\cite{Alekhin:2016jjz}~\footnote{We thank
D. Hirschb\"uhl and S. Moch for clarifications on this point.}.

\section{conclusion}

After at least thirty years of computations aiming to describe the production of a single top quark in hadronic collisions,
a lot of progress has been done.
Thanks to the relative simplicity of the hard process and the large amount of available data, the theoretical description of single top production
at hadron colliders is an excellent laboratory to study new ideas on how to treat the radiation. At the same time
in the near future, single top observables could be competitive on pdf studies and top mass determinations
and so trigger further developments.


\begin{thebibliography}{99}


\bibitem{Willenbrock:1986cr}
  S.~S.~D.~Willenbrock and D.~A.~Dicus,
  Phys.\ Rev.\ D {\bf 34}, 155, (1986).

\bibitem{Berger:2016oht}
  E.~L.~Berger, J.~Gao, C.-P.~Yuan and H.~X.~Zhu,
  Phys.\ Rev.\ D {\bf 94} no.7,  071501, (2016)

\bibitem{Jezo:2015aia}
  T.~Je\v{z}o and P.~Nason,
  JHEP {\bf 1512} 065, (2015) 

\bibitem{Frederix:2016rdc}
  R.~Frederix, S.~Frixione, A.~S.~Papanastasiou, S.~Prestel and P.~Torrielli,
  JHEP {\bf 1606} 027, (2016) 

\bibitem{Jezo:2016ujg}
  T.~Je\v{z}o, J.~M.~Lindert, P.~Nason, C.~Oleari and S.~Pozzorini,
  Eur.\ Phys.\ J.\ C {\bf 76} no.12,  691, (2016) 

\bibitem{Demartin:2016axk}
  F.~Demartin, B.~Maier, F.~Maltoni, K.~Mawatari and M.~Zaro,
  arXiv:1607.05862 [hep-ph].

\bibitem{2006}
N.~Kidonakis, Phys.\ Rev.\ D {\bf 74}, 114012, (2006).

\bibitem{2007}
N.~Kidonakis, Phys.\ Rev.\ D {\bf 75}, 071501, (2007).

\bibitem{2010}
N.~Kidonakis, Phys.\ Rev.\ D {\bf 81}, 054028, (2010).

\bibitem{Alekhin:2016jjz}
  S.~Alekhin, S.~Moch and S.~Thier,
  Phys.\ Lett.\ B {\bf 763} 341, (2016)


\end{thebibliography}
\end{document}